\documentclass[twoside,slac_one]{revtex4}
\usepackage{graphicx}
\usepackage{fancyhdr}
\usepackage{amsmath} % American Mathematics Society standards
\usepackage{bm}% bold math
\usepackage{amsxtra}
\usepackage{amssymb}
\usepackage{amsthm}
\usepackage{latexsym}
\usepackage{lscape}
\usepackage{atlasphysics}
\usepackage{xspace}

\newcommand{\sqR}   {\tilde{q}^{}_{R}}
\newcommand{\sqL}   {\tilde{q}^{}_{L}}

\newcommand{\bone}   {\tilde{b}^{}_{1}}
\newcommand{\tone}   {\tilde{t}^{}_{1}}
\newcommand{\neut}  {\tilde{\chi}^{0}_{1}}  
 
\def\etmiss{\ensuremath{E_{\mathrm{T}}^{\mathrm{miss}}}\xspace}

\def\dphimin{\ensuremath{\Delta\phi_{min}}}

\def\bjets{$b$-jets\xspace}
\def\bjet{$b$-jet\xspace}
\def\cjets{$c$-jets\xspace}

\def\btagging{$b$-tagging\xspace}
\def\btag{\ensuremath{b}-tag\xspace}
\def\btagged{$b$-tagged\xspace}

\def\meff{\ensuremath{\mathrm{m}_{\mathrm{eff}}}\xspace}
\def\mt{\ensuremath{\mathrm{m}_{\mathrm{T}}}\xspace}

\newcommand{\gl}   {\tilde{g}}
\newcommand{\sq}   {\tilde{q}}

\renewcommand{\ttbar} {\ensuremath{t\bar{t}}\xspace}
\renewcommand{\met} {\ensuremath{E_{\mathrm{T}}^{\mathrm{miss}}}\xspace}
\renewcommand{\pt} {\ensuremath{p_\mathrm{T}}\xspace}

\newcommand{\lumi}{0.83}

\begin{document}

\title{Search for New Physics at $\sqrt{s}$ = 7 TeV in Hadronic Final States with Missing Transverse Energy and Heavy Flavor}


\author{Bart Butler on behalf of the ATLAS Collaboration}
\affiliation{SLAC National Accelerator Laboratory, Menlo Park, CA, USA}

\begin{abstract}
A search for supersymmetric particles in events with large missing transverse 
momentum, heavy flavor jet candidates and no leptons ($e$,$\mu$) in $\sqrt{s} = 7$~TeV 
proton-proton collisions is presented. In a data sample corresponding to an integrated luminosity of 0.83 ~fb$^{-1}$ recorded by the 
ATLAS experiment at the Large Hadron Collider, no significant excess is observed with respect to the prediction for Standard Model processes.
Model-independent production cross section upper limits are provided in the context of simplified 
models as well as conventional limits.
\end{abstract}

\maketitle

\thispagestyle{fancy}


%%%%%%%%%%%%%%%%%%%%%%%%%%%%%%%%%%
\section{Introduction}
Supersymmetry (SUSY)~\cite{Golfand:1971iw} is one of the most compelling theories 
to describe physics beyond the Standard Model (SM). In the framework of a generic $R$-parity conserving minimal 
supersymmetric extension of the SM, the MSSM~\cite{Martin:1997ns}, SUSY particles are produced in pairs and 
the lightest supersymmetric particle (LSP) is stable. In a large variety of models, the LSP 
is the lightest neutralino, $\neut$, which is weakly interacting and is 
a possible candidate for dark matter. 
The coloured superpartners of quarks and gluons, the squarks ($\sq$) and 
gluinos ($\gl$), are expected to be copiously produced via the strong 
interaction at the Large Hadron Collider (LHC).
The partners of the right-handed and left-handed quarks, $\sqR$ and $\sqL$, 
can mix to form two mass eigenstates. These mixing effects are proportional to 
the corresponding fermion masses and therefore become important for the third 
generation. In particular, large mixing can yield sbottom ($\bone$) and stop ($\tone$) 
mass eigenstates that are significantly lighter than other squarks. 
Consequently, $\bone$ and $\tone$ could be produced with large cross sections at 
the LHC, either via direct pair production or, if kinematically allowed, 
through $\gl\gl$ production with subsequent $\gl \rightarrow \bone b$ or  $\gl \to \tone t$ decays.
Depending on the SUSY particle mass spectrum, the cascade decays of gluino-mediated and 
pair-produced sbottoms or stops result in complex final states consisting of 
missing transverse momentum (its magnitude is referred to as 
$\etmiss$ in the following) and several jets, among which $b$-quark jets ($b$-jets) 
are expected. 

SUSY is searched for in final states involving $\etmiss$, 
energetic jets, of which at least one must be identified as a $b$-jet and no 
isolated leptons ($e$ or $\mu$). The search is based on $pp$ collision 
data at a centre-of-mass energy of 7~TeV recorded by the ATLAS experiment~\cite{DetectorPaper:2008} at the 
LHC in 2011. The total data set included in the analysis corresponds to an integrated luminosity of 
\lumi~$\ifb$.

Two phenomenological 
MSSM scenarios are considered where the first and second generation squark masses are set 
above 2 TeV. In the first scenario, the $\bone$ is the lightest squark,
$m^{}_{\gl}>m^{}_{\bone}>m^{}_{\neut}$, and the branching ratio 
for $\gl \rightarrow \bone b$ decays is 100\%. Sbottoms are produced 
via gluino-mediated processes or via direct pair production and they 
are assumed to decay exclusively via $\bone \rightarrow b\neut$, where 
$m^{}_{\neut}$ is fixed at 60~GeV. The interpretation of the results is presented as a function of the gluino and light sbottom masses. 
The second MSSM-like scenario is defined in the context of the general 
simplified models~\cite{Alves:2011wf}: all squarks including $\bone$ are heavy, 
gluino-pair production is the only kinematically allowed process and 
gluinos decay (off-shell) into~$b\bar{b}\neut$ final states. Here the results are interpreted in a  ($m_{\gl},m_{\neut}$) plane. 
These results are generalized to any new physics process where  
gluino-like particles decay into $b\bar{b}$ and a weakly interacting 
massive particle.

%%%%%%%%%%%%%%%%%%%%%%%%%%%%%%%%%%
\section{Monte Carlo Simulated Samples}\label{sec:SimEvSamp}
Simulated event samples are used to determine the detector acceptance, the 
reconstruction efficiencies and the expected event yields for signal and 
background  processes. Samples of SUSY signal processes were simulated for various models using the 
{\tt HERWIG++}~\cite{herwig} v2.4.2 Monte Carlo program. 
The particle mass spectra and decay modes were determined using the {\tt SUSYHIT}~\cite{susyhit} v1.3 
program, and parametrized 
in the ($m^{}_{\gl},m^{}_{\bone}$) and ($m^{}_{\gl},m^{}_{\tone}$) planes. 
The SUSY sample yields are then normalized to the expectations from 
next-to-leading order (NLO) calculations obtained using 
the {\tt PROSPINO}~\cite{prospino} v2.1 program. 
For these calculations the CTEQ6.6M~\cite{cteq6m} parametrisation of the parton density functions (PDFs)
is  used and the renormalization and factorisation scales are set to the 
average mass of the sparticles produced in the hard interaction.   

\begin{table}[t]
\begin{small}
\begin{center}
\begin{tabular}{ l|r r}
\hline
\hline
\raisebox{-0.4ex}{Physics process} &  {\raisebox{-0.4ex}{$\sigma \cdot$ BR [nb]}} &  \\
\hline
\hline
$W \rightarrow \ell \nu$ (+jets) & 31.4$\pm$1.6  &\cite{Hamberg:1990np,Melnikov:2006kv,Melnikov:2006di}  \\
$Z/\gamma^* \to \ell \ell$ (+jets) & 3.20$\pm$0.16 &\cite{Hamberg:1990np,Melnikov:2006kv,Melnikov:2006di} \\

$Z \to \nu \nu$ (+jets)  & 5.82$\pm$0.29 &\cite{Hamberg:1990np,Melnikov:2006kv,Melnikov:2006di} \\

$\ttbar$          & 0.165$^{+0.011}_{-0.016}$  &\cite{Bonciani:1998vc,Moch:2008qy,Beneke:2009ye}\\
Single top         & 0.085$\pm$0.003  &\cite{Kidonakis:2011wy,Kidonakis:2010tc}\\
\hline
\hline
\end{tabular}
\caption{The most important background processes and their predicted cross sections, 
multiplied by the relevant branching ratios (BR) before any event selection. A generator level cut $m_{\ell
\ell}>40$~GeV was
applied to the $Z/\gamma^*(\rightarrow \ell^+ \ell^-)$ process.  Contributions from higher 
order 
QCD corrections are included for  
$W$ and $Z$ boson production (NNLO corrections) and for \ttbar\ production 
(NLO+NNLL corrections).}
\label{MC}
\end{center}
\end{small}
\end{table}

For the background, the following Standard Model processes are considered: 
\begin{itemize} 
\item $\ttbar$ and single top production: events were generated using the generator 
{\tt MC@NLO}~\cite{mcatnlo,mcatnlo2} v3.41. For the evaluation of systematic uncertainties, additional 
$\ttbar$ samples were generated using the {\tt POWHEG}~\cite{powheg}, {\tt ALPGEN}~\cite{alpgen} and {\tt ACERMC}~\cite{Kersevan:2004yg} 
programs. 

\item $W (\rightarrow \ell \nu)$+jets, $Z/\gamma^*(\rightarrow \ell^+ \ell^-)$+jets  
(where $\ell=e,\ \mu,\ \tau$) and 
$Z(\rightarrow \nu \bar{\nu})$ +jets  
production: events with light and heavy ($b$,$c$) flavor jets were generated using the 
{\tt ALPGEN} v2.13 program. A generator level cut $m_{\ell 
\ell}>40$~GeV was 
applied to the $Z/\gamma^*(\rightarrow \ell^+ \ell^-)$ process.

\item Di-boson ($WW$, $WZ$ and $ZZ$) production: events were generated using {\tt ALPGEN}, 
however, compared to the other backgrounds their
contribution was found to be negligible, after the application of the selection criteria. 
\end {itemize} 

\noindent For the QCD background, no reliable prediction can be obtained from a leading order 
Monte Carlo simulation and a data-driven method is used to determine the contribution 
 to the selected event samples, as discussed in Section~\ref{sec:wbkg}.

All signal and background samples were generated at $\sqrt{s}=7$~TeV using the ATLAS 
MC10 parameter tune~\cite{mc10tunes}, processed with the {\tt GEANT4}~\cite{geant4}~simulation 
of the ATLAS detector~\cite{atlassimulation}, then reconstructed 
and passed through the same analysis chain as the data. For all generators, except for 
{\tt PYTHIA}, the {\tt HERWIG + JIMMY}~\cite{herwig, Jimmy} modelling of the parton shower and 
underlying event was used (v6.510 and v4.31, respectively). 

For the comparison to data, all non-QCD background cross sections 
are normalized to the results of higher order QCD calculations. 
A summary of the relevant cross sections is given in Table~\ref{MC}.
For the next-to-next-to-leading order (NNLO) $W$ and $Z/\gamma^*$ production cross sections, an uncertainty of $\pm$5\% is 
assumed. For the $\ttbar$ production cross section, the 
corresponding uncertainty on the NLO+NNLL (next-to-next-to-leading logarithms) cross section is  estimated to be 
$^{+6.5\%}_{-9.5\%}$.

%The LHC pile-up conditions are taken into account by reweighting events according to the mean number of interactions
%expected.

All Monte Carlo samples are generated with both in-time and out-of-time pile-up from multiple proton--proton interactions. 
The simulated events are reweighted such that the distribution of interactions per 
crossing in the Monte Carlo matches the one observed in data.

%%%%%%%%%%%%%%%%%%%%%%%%%%%%%%%%%%
\section{Data and Baseline Event Selection}
After the application of beam, detector and 
data-quality requirements, the data set used for this analysis correspond to a total integrated luminosity of \lumi\ \ifb. 

Events are selected at the trigger level by 
requiring one jet with high $\pt$ and large missing transverse momentum. The selection is fully efficient 
for events containing at least one jet with $\pt >130$~GeV and $\etmiss>130$~GeV~\cite{MissingEtTrigger}. 

In the data sample selected, jet candidates are reconstructed using the anti-$k_t$ jet 
clustering algorithm~\cite{anti-kt, Cacciari200657, Fastjet} with a distance parameter of $R$=0.4. The inputs to this 
algorithm are three dimensional calorimeter energy clusters. 
The reconstructed jet energies are corrected for inhomogeneities and for the non-compensating nature 
of the calorimeter by using $\pt$- and $\eta$-dependent 
calibration factors determined from Monte Carlo simulation and 
validated using extensive test-beam measurements and studies of $pp$ collision data (Ref.~\cite{JES} and references therein). 
Only jets with $\pt>20$~GeV and within $|\eta|<2.8$ are retained for this analysis. 
Candidates for $b$-jets are identified among jets with $\pt>50$~GeV using 
an algorithm that reconstructs a  
vertex from all tracks which are displaced from the primary vertex and associated with the jet. 
The parameters of the algorithm are chosen such that a tagging efficiency of 50\% (1\%) is 
achieved for $b$-jets (light flavor or gluon jets) in $\ttbar$ events in Monte Carlo simulation~\cite{btagEffCal}. 

In order to apply the lepton veto, electron candidates are required to 
satisfy the `medium'  selection criteria, as detailed in  
Ref.~\cite{elrecon}. 
Muon candidates are identified either as a match between an extrapolated inner 
detector track and one or more segments in the muon spectrometer, or by associating 
an inner detector track to a muon spectrometer track. The combined track parameters 
are derived from a statistical combination of the two sets of track parameters.
Electrons (muons) are required to have $\pt>20$ GeV (10 GeV) and $|\eta|<2.47 (2.4)$. 

The calculation of $\etmiss$ is based on the
modulus of the vectorial sum of the $\pt$ of the reconstructed jets
(with $\pt > 20 \GeV$ and over the full calorimeter coverage $|\eta| < 4.9$), leptons 
  (including non--isolated muons) and the
calorimeter clusters not belonging to reconstructed objects.

After object identification, overlaps are resolved.
Any jet within a distance $\Delta R $ = 0.2
of a medium electron candidate is discarded and any remaining lepton within $\Delta R=$ 0.4 of a jet is discarded. 

Events are selected if the primary vertex is associated with five or more tracks. They   
are also required to pass basic quality criteria to discriminate against detector noise 
and non-collision backgrounds. 
Due to a front--end electronics failure  in one of the electromagnetic calorimeter modules, a region of the calorimeter 
of size $\Delta \eta \times \Delta \phi = 1.6 \times 0.4$ was only partially read out.  Events with any jet with $\pt > 
50$ GeV in this region are rejected. The acceptance loss caused by this selection cut is about 10\%.

Selected events are required to have at least one jet with 
$\pt>130$~GeV, at least two additional jets with $\pt>50$~GeV and $\etmiss>130$~GeV. At least one jet is required to be $b$-tagged. 
Events containing electron or muon candidates are rejected.   
The effective mass, $\meff$, is defined as the scalar sum of $\etmiss$ and the 
transverse momenta of the three leading jets. Events are required to 
have $\etmiss / \meff>0.25$. In addition, the smallest azimuthal separation between the 
$\etmiss$ direction and the three leading jets, $\dphimin$, is required to be larger 
than 0.4. 
The last requirement effectively reduces the amount of QCD  background  where $\etmiss$ results from mis-reconstructed jets or from 
neutrinos emitted along the direction of the jet axis by heavy flavor decays.

%%%%%%%%%%%%%%%%%%%%%%%%%%%%%%%%%%
\section{Signal Region Optimization}

Phenomenological models have many advantages as signal samples when attempting to design analyses sensitive to as broad a slice of phase space as is both possible and practical. They describe well-motivated, SM-like production and decay processes, and the kinematics are determined by a small number of parameters (masses). Both SUSY scenarios considered in this analysis result in 4 $b$-jet + $\etmiss$ final state signatures. The simplified model kinematics are determined to first order by a single parameter: the mass difference between the gluino and the neutralino, $\Delta M(\gl-\neut)$. This simplicity motivates the choice of this model for signal region optimization studies, with cross-checks performed to ensure the results are relevant in the more complicated case.

The optimization procedure was designed to ensure broad sensitivity given the baseline selection constraints. Several likely kinematic variables were evaluated for signal/background separation power: $\etmiss$, $\meff$, jet multiplicity, b-tagged jet multiplicity, etc. From this, the best variable set was chosen and an n-dimensional cut grid was created. For each  set of cuts and simulated signal sample, a systematic-corrected significance was computed (systematics estimated from Monte Carlo) and used to assemble a set of optimal cuts for maximal sensitivity to the entire ($m_{\gl},m_{\neut}$) plane.

This list of "optimal" signal regions included nearly unique sets of cuts for each mass point. The number of selections were gradually reduced while ensuring broad-based sensitivity was retained. Four signal regions were chosen to represent good compromises among mass plane coverage, sensitivity, and practical concerns such as background control regions. They are characterized by the minimum number of 
$b$-jets required in the final state and by the threshold of the further 
selection on \meff: 3JA ($\geq$1 \bjet, \meff$>500$~GeV),  3JB ($\geq$1 \bjet, \meff$>700$~GeV),  
3JC ($\geq$2 \bjet, \meff$>500$~GeV) and 3JD ($\geq$2 \bjet, \meff$>700$~GeV). 

%%%%%%%%%%%%%%%%%%%%%%%%%%%%%%%%%%
\section{Standard Model Background Estimation}\label{sec:wbkg}
The expected  amount of  \ttbar, $W/Z$+jets and single top events is estimated using the Monte Carlo 
simulation. Events from \ttbar production represent the largest background component in all four signal regions. The Monte Carlo 
prediction is validated by a data-driven estimate which relies on control regions with the same kinematic 
selection on jets and missing transverse momentum and an electron or a
muon with \pt$>20$~GeV in the final state, \meff$>600$~GeV and $40$~GeV$< \mt < $100 GeV (where $\mt$
is the transverse mass computed from the lepton 4-vector and the \etmiss) and at least one or two $b$-jets. The normalization 
determined in these control regions (corrected for non-$\ttbar$ contamination) is then transferred to the kinematically similar signal regions. 
Figure~\ref{fig:1lCR} shows the $\meff$ distribution in the 1-electron and 1-muon (1 $b$-tag) control regions. 
The agreement between the data and the Monte Carlo prediction is good and, as a consequence, the prediction of the 
data-driven estimation agrees with that of the Monte Carlo.

\begin{figure}[h!]
  \begin{center}
    $\begin{array}{c@{\hspace{.1in}}c}
      \includegraphics[width=0.45\columnwidth]{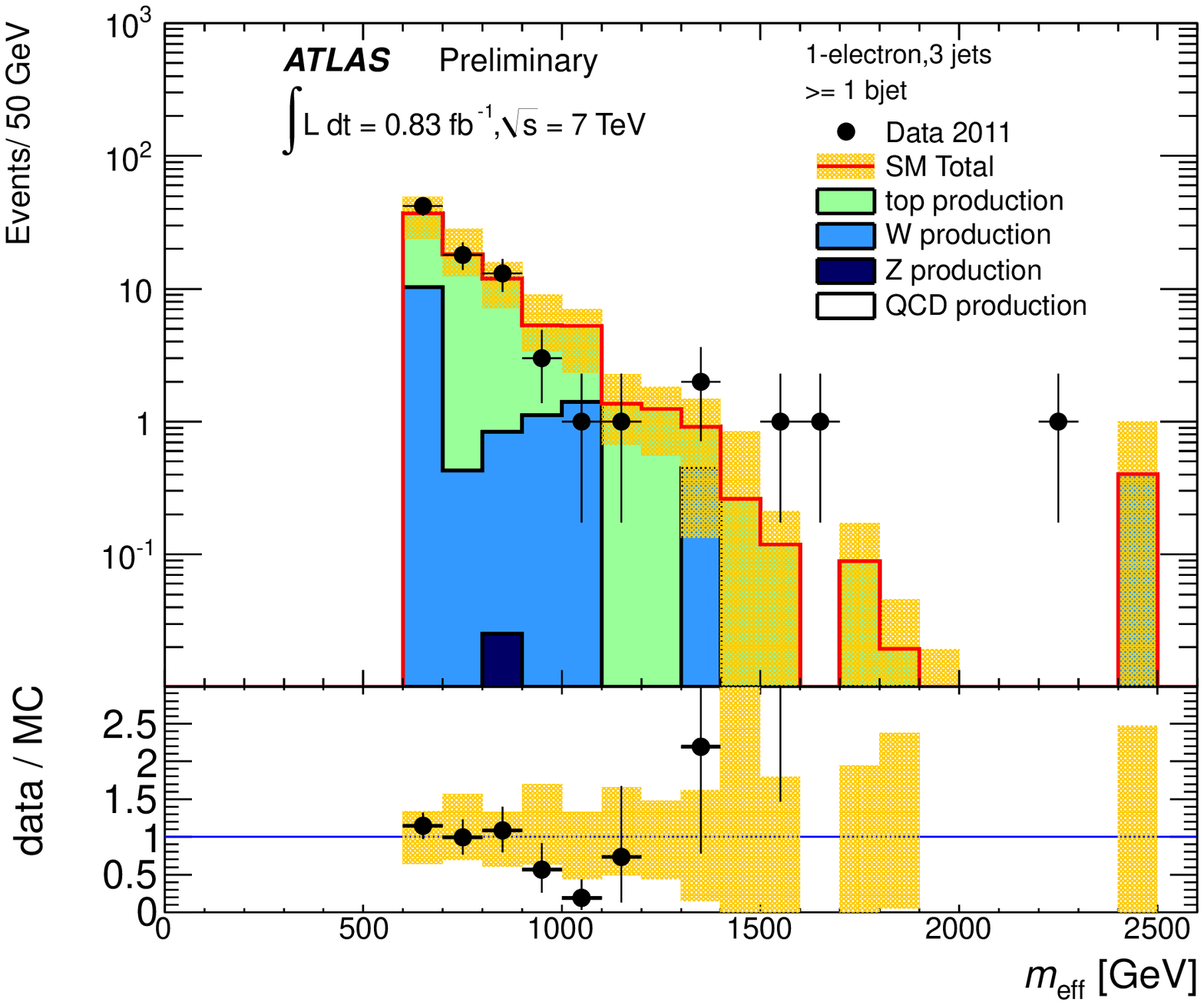} &
      \includegraphics[width=0.45\columnwidth]{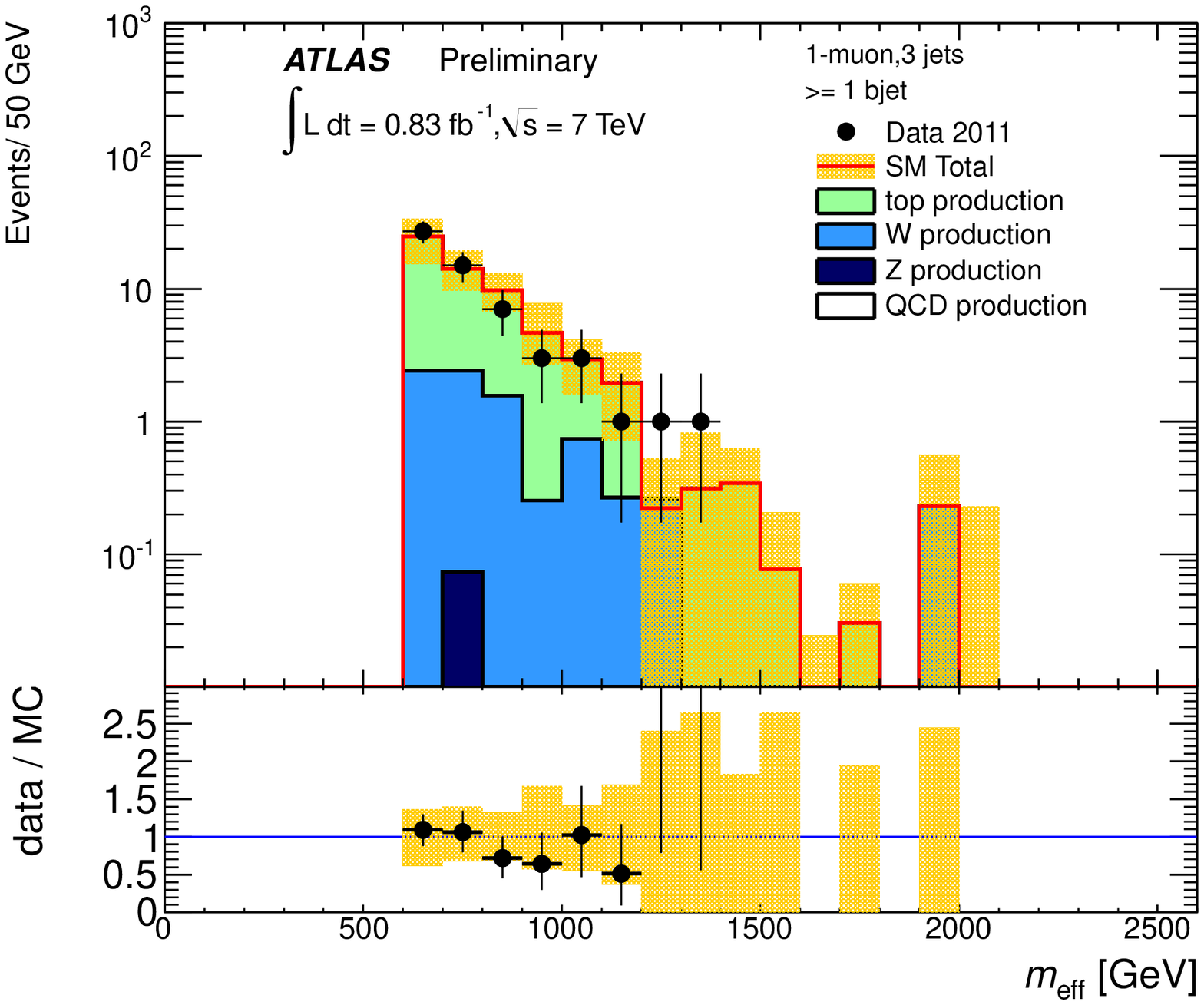} \\
      \end{array}$
    \end{center}
  \caption{Effective mass distribution for the 1-electron (left) and 1-muon (right) \ttbar control regions. The lower plot shows the bin-by-bin ratio of the data to the MC. The yellow band shows the full systematic uncertainty on the SM expectation.}
  \label{fig:1lCR}
\end{figure}

The Monte Carlo estimation of the $W/Z$ background, which is done using {\tt ALPGEN} as baseline generator,  includes 
a dedicated simulation of $W +$ heavy flavor quarks. Double counting arising both at generator and at Matrix 
Element/Parton Shower matching level between samples generated with light and heavy flavor quarks is resolved using a 
$\Delta R$ matching between partons and jets and a dedicated overall normalization scale factor 
% of $1.63 \pm 0.76$~\cite{HFnorm} 
is applied to the $W +$ heavy flavor samples. For each signal region, the normalization of the inclusive $W/Z$ Monte Carlo 
prediction is 
validated with a combined fit of $\ttbar$ and $W/Z$ background components to the distribution of the number of $b$-tagged jets in a 0-lepton control region defined by reverting the $\meff$ cut. The fit confirms the Monte Carlo prediction.  

Since its contribution to the total background is small, the estimation for the single top background is based entirely on the Monte Carlo prediction.

The remaining QCD background in the signal regions is estimated with a data driven procedure. 
The technique~\cite{daCosta:2011qk,ATLAS-CONF-2011-086} used is to smear the momentum of jets in clean data 
events with low $\etmiss$ to generate "pseudoevents" with possibly large $\etmiss$
values.
The method was validated by comparing data and pseudoevents distributions in QCD enriched control 
regions that are kinematically similar to the signal regions,  obtained by reverting the cut 
on $\dphimin$. Figure~\ref{fig:QCDCR} shows the \meff distribution for events with $\dphimin < 0.4$ 
and 1 (left) or 2 (right) $b$-tagged jets. The emulated QCD distributions agree with the data.

\begin{figure}[h!]
  \begin{center}
    $\begin{array}{c@{\hspace{.1in}}c}
      \includegraphics[width=0.45\columnwidth]{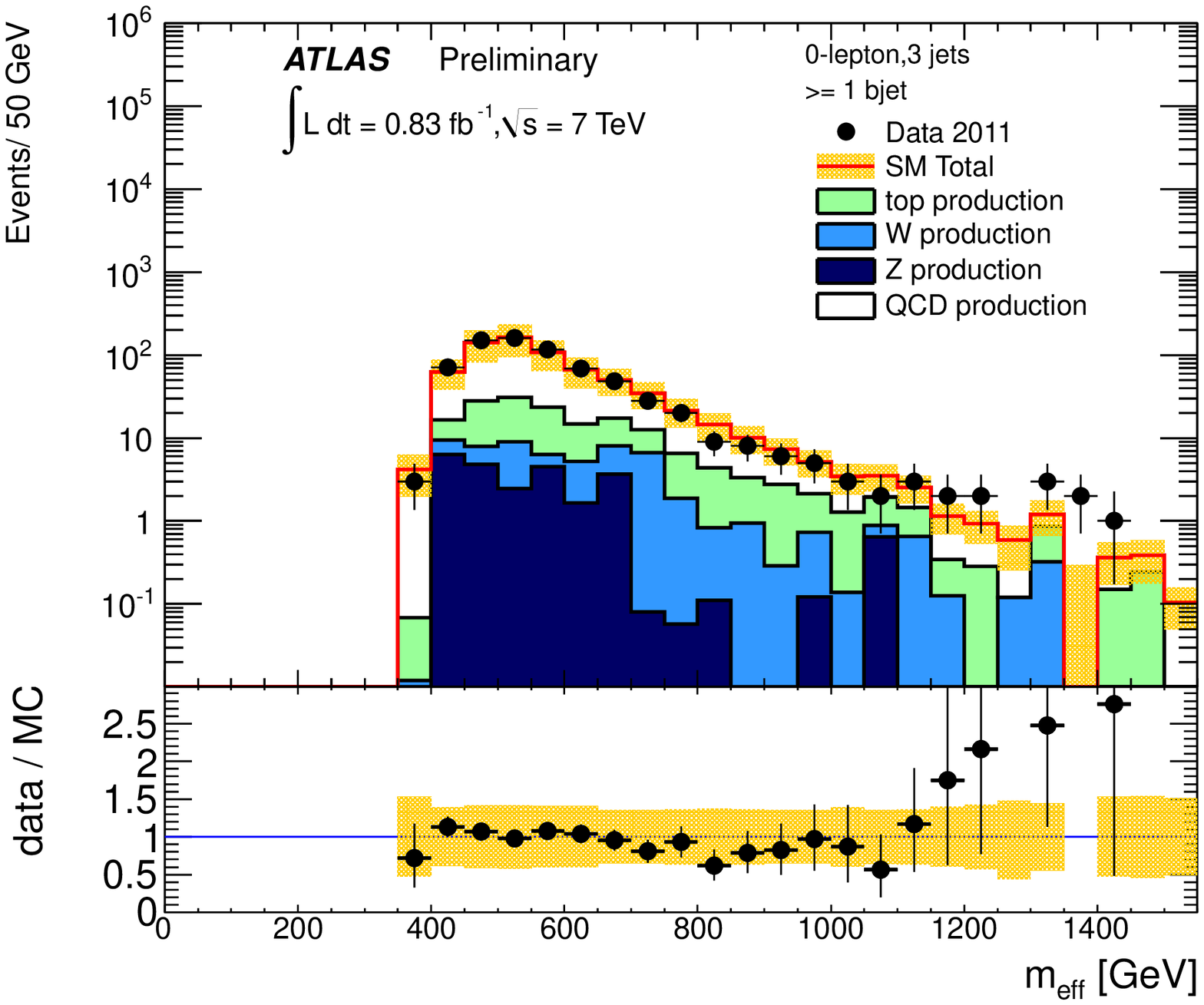} &
      \includegraphics[width=0.45\columnwidth]{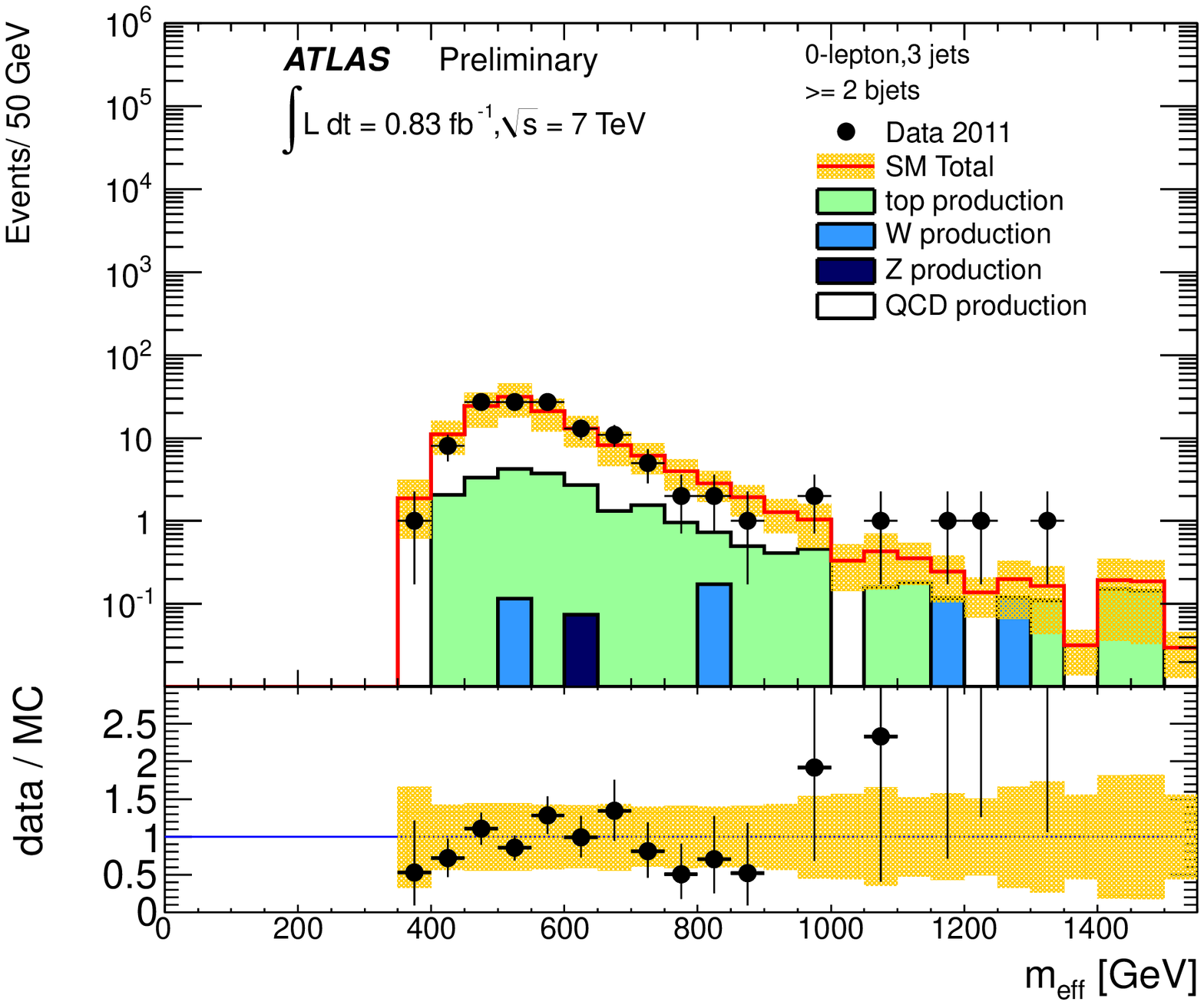} \\
      \end{array}$
    \end{center}
  \caption{Effective mass distribution for the QCD control region ($\dphimin < 0.4$) for events with 1 (left) or 2 (right) $b$-tagged jets. The lower plot shows the bin-by-bin ratio of the data to the SM expectation. The QCD prediction validated in these control samples is derived using the smearing method described in the text, other SM background contributions are estimated from MC. The yellow band shows the full systematic uncertainty on the SM expectation }
  \label{fig:QCDCR}
\end{figure}

%%%%%%%%%%%%%%%%%%%%%%%%%%%%%%%%%%
\section{Systematic Uncertainties}\label{syst}
The background from top and $W/Z$ production is
obtained using the Monte Carlo simulation. The total uncertainty on this prediction 
is estimated to be between $\pm$30\% and $\pm$35\% depending on the final selection. 
It is dominated by the uncertainty on the jet energy scale, on the theoretical 
prediction of the background processes and on the determination of 
the \btagging efficiency. %~\cite{btagEffCal}. 
The uncertainty on the jet energy scale (derived using 2010 collision data~\cite{JES}) varies as a function of the jet \pt and pseudorapidity 
and it is about 2\% at $\pt = 50$ GeV in the central detector region.
Additional systematic uncertainties arise from the dependence of the jet response  on the number of 
expected interactions per bunch crossing and on the jet flavor. 
The total jet energy scale uncertainty at 50 GeV in the central detector region is about 5\%.
This translates into a 20--25\% uncertainty on the absolute prediction of the 
background from SM processes.  
Uncertainties on the theoretical cross sections of the background processes (see Section~\ref{sec:SimEvSamp}), 
on the modelling of initial and final-state soft gluon radiation and the limited knowledge of the 
PDFs of the proton lead to uncertainties of $\pm$25\% and $\pm$30\% on the absolute predictions of the \ttbar\ and  
the $W/Z$+jet backgrounds, respectively.  An additional uncertainty of 50\%(100\%) is assigned 
to the associated production of $W(Z)$ and heavy flavor jets. The uncertainty on the determination 
of the tagging efficiency for \bjets, \cjets and light flavor jets introduces further 
uncertainties on the predicted background contributions at the level of 
$\pm$10\% ($\pm$22\%) for \ttbar\ and 
$\pm$15\% ($\pm$30\%) for $W/Z$+jets in the 1 (2) $b$-tag signal regions. 
For the QCD background estimation, the uncertainty of 50\% is dominated by the dependency 
of the smearing function on the flavor composition of the low \etmiss sample used as 
smearing starting point. 
%For the QCD  background estimation, the uncertainty of 50\% is dominated by the dependence of the smearing function on the flavor composition of the seed events used for the data driven estimation.

For the SUSY signal processes, various sources of uncertainties 
affect the theoretical NLO cross sections.
Variations of the renormalization and factorisation scales by a factor 
of two result in uncertainties of $\pm$16\% for 
$\gl\gl$ production and $\pm$30\% for $\bone\bone$ pair 
production, with little dependence on the sparticle masses and the 
SUSY model. 

The number of predicted signal events is also affected by the 
PDF uncertainties, estimated using the CTEQ6.6M PDF error 
eigenvector sets at the 90\% C.L. limit, rescaled to 1$\sigma$. 
The relative uncertainties on the $\gl\gl$ ($\bone\bone$) cross sections 
were estimated to be in the range from $\pm$11\% to $\pm 25\%$ 
($\pm 7\%$ to $\pm 16\%$) for the $\gl\gl$ ($\bone\bone$) processes, 
depending on the gluino and sbottom masses. 
Uncertainties due to the modelling of initial and final state radiation on the signal are not included. 
 
The impact of detector-related uncertainties, such as the jet energy scale (JES) 
and \btagging uncertainties, on the signal event yields depends on the masses of the  produced sparticles. The total uncertainty varies between 
$\pm$20\% ($\pm 35\%$)  and $\pm$10\% ($\pm 10\%$) for the 1 (2) $b$-tag case  
as the gluino/sbottom masses increase from 200~GeV to 1~TeV, across the 
different scenarios, and it is dominated by the JES and the \btagging 
uncertainty for low and high mass sparticles, respectively. 

Finally, an additional $\pm$4.5\% uncertainty on the quoted total integrated 
luminosity was taken into account, based on the 2010 luminosity calibration~\cite{EPSLumi2011} 
which was transferred to the 2011 data by using the LAr forward calorimeter and the 
tile calorimeter current measurements.

%%%%%%%%%%%%%%%%%%%%%%%%%%%%%%%%%%
\section{Results}
Good agreement between data and Monte Carlo prediction is observed in the distributions of the $\meff$, 
the $\met$ and the \pt of the leading jet as shown in  Figure~\ref{fig:datamc3jSRs} before the 
\meff cut for the signal regions with one and two $b$-tags.  
\begin{figure}
  \centering
\includegraphics[width=0.45\columnwidth]{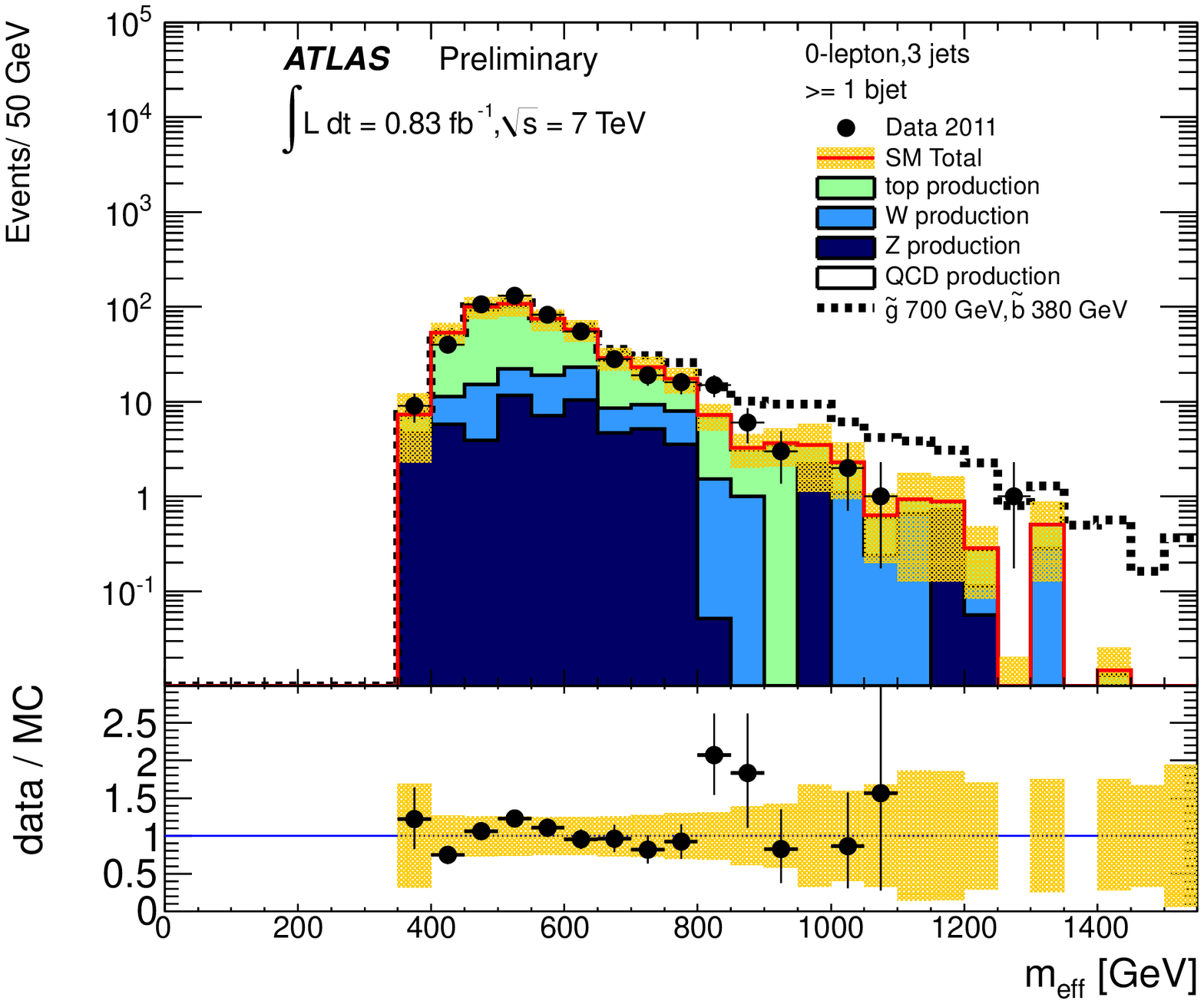}
\includegraphics[width=0.45\columnwidth]{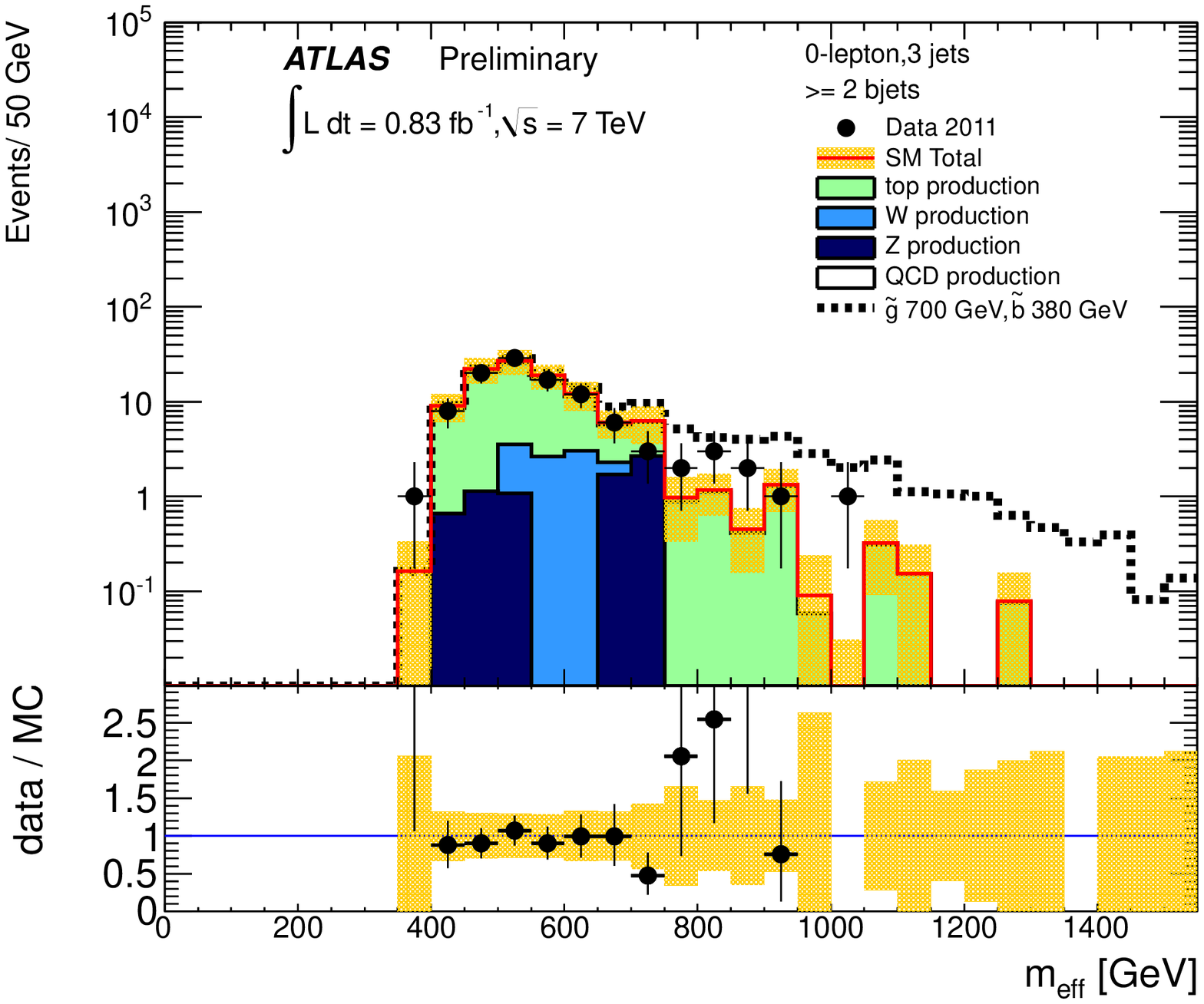}
\includegraphics[width=0.45\columnwidth]{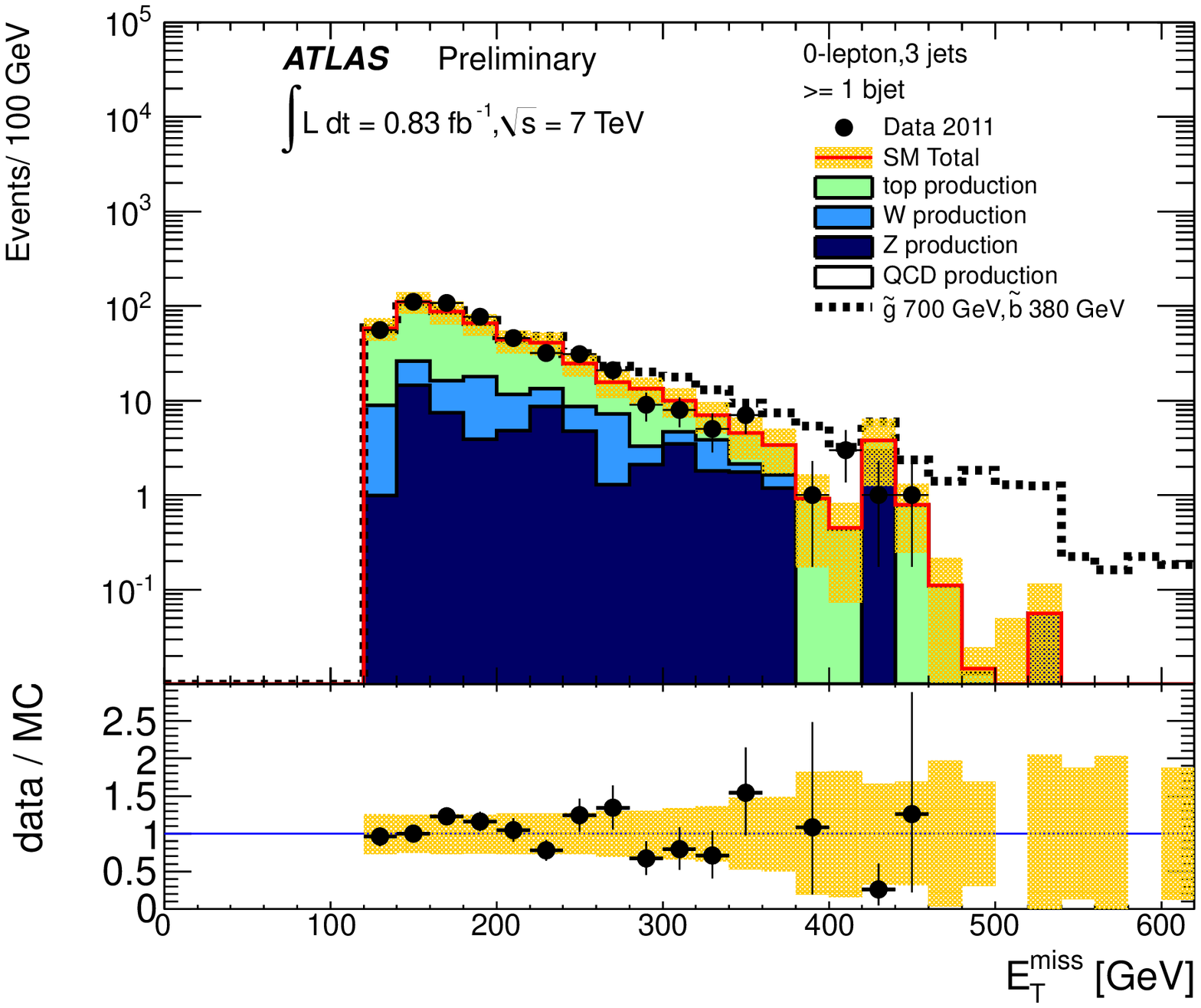}
\includegraphics[width=0.45\columnwidth]{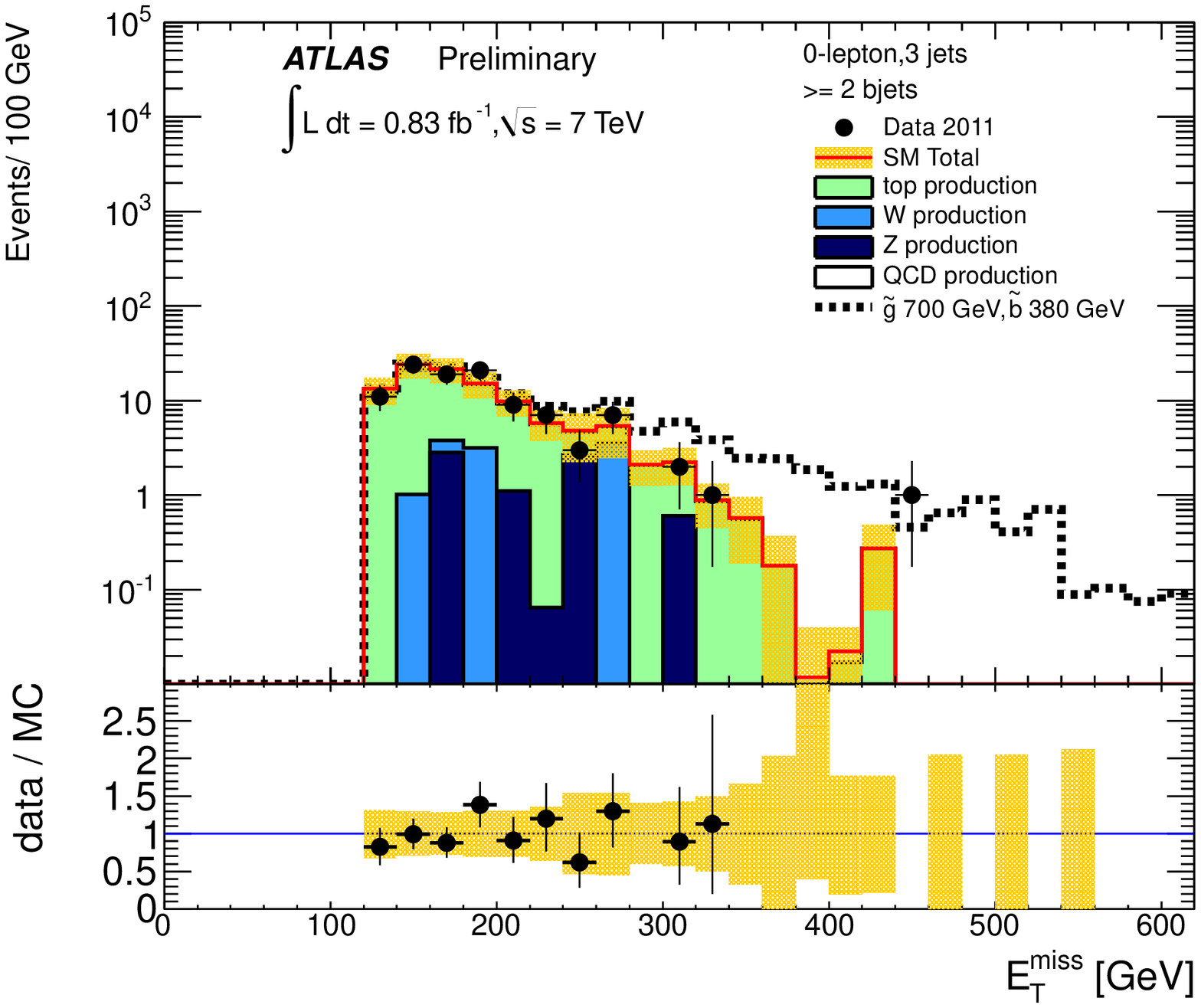}
\includegraphics[width=0.45\columnwidth]{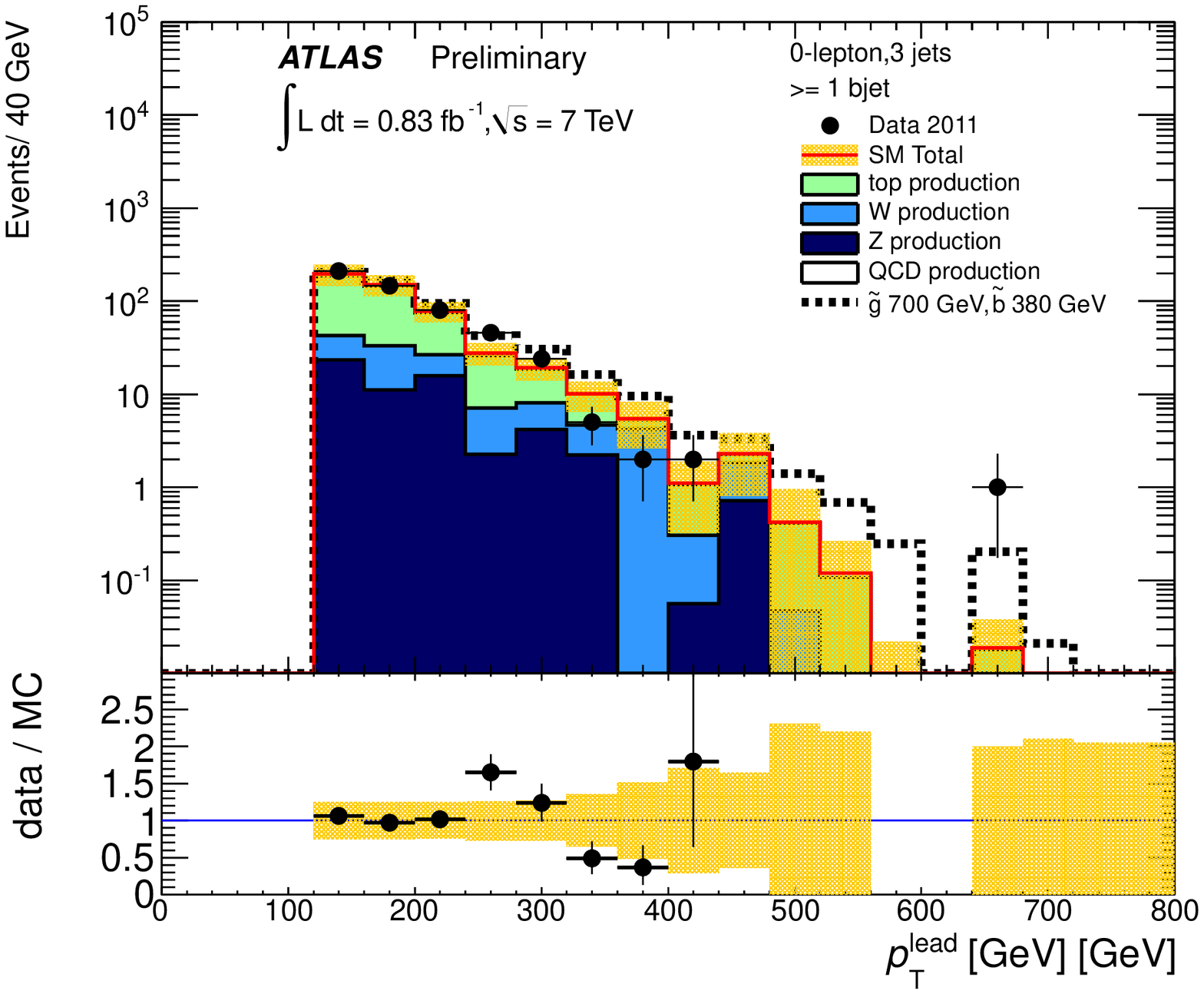}
\includegraphics[width=0.45\columnwidth]{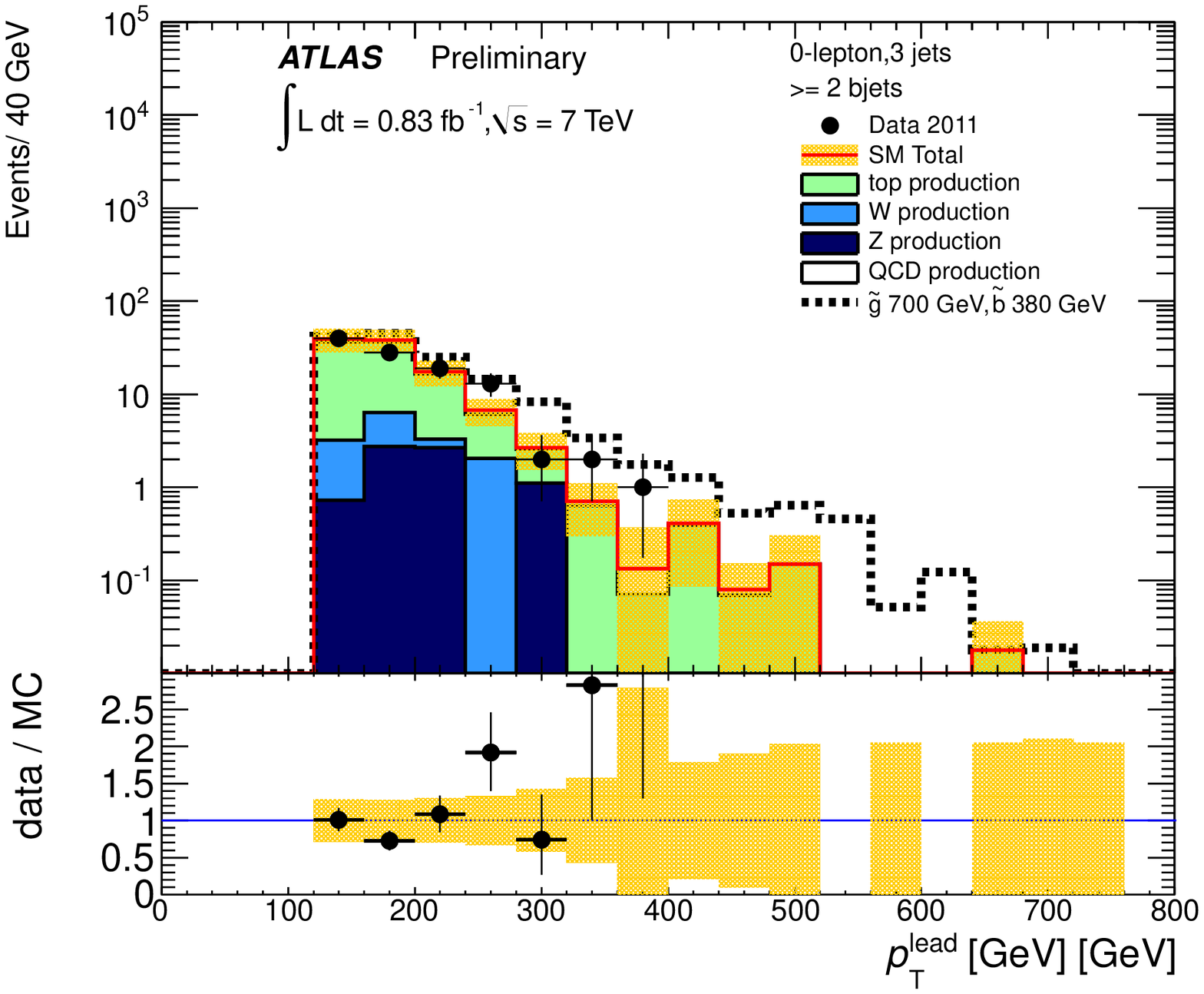}
  \caption{Distributions of the effective mass, $\meff$ (top), $\met$ (middle) and the \pt 
of the leading jet (bottom) for data and the expected SM processes in the 1 $b$-tag (left) 
and 2 $b$-tags (right) signal regions.  The yellow band shows 
the full systematic uncertainty on the SM expectation. For illustration, the distributions 
of one reference SUSY signal are superimposed. The lower plot shows the bin-by-bin ratio 
of the data to the SM expectation.}\label{fig:datamc3jSRs}
\end{figure}

The observed and predicted event yields in the four signal regions are given in Table~\ref{tab:3jresults} 
together with the total estimated uncertainty on the predictions. The value used for the top background estimate includes the \ttbar and the single top contributions as predicted by the Monte Carlo. The \ttbar component is validated using a dedicated data driven procedure.  
The $W/Z$+jets background is estimated using Monte Carlo, and the uncertainties correspond to those discussed 
in Section~\ref{syst}. QCD-multijet contributions are estimated with the jet smearing method. 
The SM predictions agree with the observed number of events in all four signal regions. 
 
\begin{table}[htp]
 \centering
 \begin{tabular}{c|c|c|c|c|c}
   \hline
   \hline
   Sig. Reg. & Data (\lumi\ $\ifb$) & Top & W/Z &  QCD & Total \\
   \hline
  3JA (1 btag $\meff>$500 GeV) & 361 & $221^{+82}_{-68}$ & $121 \pm 61$ & $15 \pm 7$ & $356^{+103}_{-92}$\\
  3JB (1 btag $\meff>$700 GeV) & 63 & $37^{+15}_{-12}$ & $31 \pm 19$  & $1.9 \pm 0.9$ & $70^{+24}_{-22}$\\
  3JC (2 btag $\meff>$500 GeV) & 76 & $55^{+25}_{-22}$ & $20 \pm 12$ & $3.6 \pm 1.8$ & $79^{+28}_{-25}$\\
  3JD (2 btag $\meff>$700 GeV) & 12 & $7.8^{+3.5}_{-2.9}$ & $5 \pm 4$  & $0.5 \pm 0.3$& $13.0^{+5.6}_{-5.2}$\\
   \hline
   \hline
   \end{tabular}
   \caption{Summary observed and expected event yields in the four signal regions. 
The QCD prediction is based on the jet smearing method described in the text. Systematic 
uncertainties for the Standard Model predictions are given.}
\label{tab:3jresults}
\end{table}

Since no excess with respect to the SM predictions is observed in the data, the results 
are translated into 95\% C.L. upper limits on contributions from new physics. 
Limits are derived using the $CL_s$~\cite{Read:2002hq} method, while the power constrained limit 
(PCL)~\cite{Cowan:PCL} method is used for comparison with previous ATLAS results. 

The results are interpreted in terms of 95\% C.L. exclusion limits for the SUSY scenarios described in the introduction. 
In Figure~\ref{fig:sb_gl_obs} the observed and expected exclusion regions 
are shown in the ($m^{}_{\gl},m^{}_{\bone}$) plane for the hypothesis that the lightest 
squark $\bone$ is produced via gluino-mediated or direct pair production and 
decays exclusively via $\bone \rightarrow b\neut$. The NLO cross sections are 
calculated using {\tt PROSPINO}. For each scenario, the signal region resulting in the best expected exclusion limit is used: the selection 3JD provides  
the best sensitivity in most cases. If $\Delta M(\gl-\bone)<100$~GeV, signal regions 
with 1 \btag are preferred, due to the lower number of expected \bjets above \pt thresholds.  
The regions 3JA and 3JB provide the best sensitivity when 
$m^{}_{\gl} \gg m^{}_{\bone}$ and sbottom pair production dominates. 
All systematic uncertainties on the signal and background contributions are taken into account in these limits and include the 
fully-correlated detector-type uncertainties (JES, \btagging, trigger, pile-up effects, 
luminosity) as well as the theoretical uncertainties on the signal (renormalization/factorization 
scale and PDF). 
Gluino masses below 720~GeV are excluded for sbottom masses up to 600~GeV. The exclusion 
is less stringent in the region with low $\Delta M(\gl-\bone)$, where low \met is expected.  
This search extends the previous ATLAS exclusion limit in the same scenario 
by about 130~GeV \cite{Aad2011398} (180~GeV if using the same limit setting procedure).  
%The excluded area is compared to previous 
%results from CDF searches in similar MSSM scenarios, and to exclusion limits from the CDF and 
%D0 experiments on direct sbottom pair production as well as the ATLAS result obtained using 35 pb$^{-1}$.

Results are also interpreted in the context of simplified models. In this case, all the squarks are heavier than the gluino, which decays exclusively into three-body final states ($b\bar{b}\neut$) via an off-shell sbottom.  
Such a scenario %, defined in $m^{}_{\gl},m^{}_{\neut}$ at fixed large sbottom mass, 
can be considered complementary to the previous one. % defined in $m^{}_{\gl},m^{}_{\bone}$ at fixed $\neut$ mass. 
The exclusion limits obtained on the ($m^{}_{\gl},m^{}_{\neut}$) plane are shown in Figure~\ref{fig:GbbMaxAll} for gluino masses above 200 GeV.  
For each combination of masses, the analysis providing the best expected limit is chosen. 
The selection 3JD leads to the best sensitivity for gluino masses above 400 GeV 
and $\Delta M(\gl-\neut)>100$~GeV. At low $\Delta M(\gl-\neut)$, soft \bjets 
spectra and low \etmiss are expected, giving higher sensitivity to the signal regions 3JA and 3JB are preferred. 
Low gluino mass scenarios present moderate \meff and high \bjet multiplicity, thus 
favouring signal region 3JC. Neutralino masses below 200-250~GeV are 
excluded for gluino masses in the range 200-660~GeV, if $\Delta M (\gl-\neut)>$100 GeV.

The results can be generalized in terms of 95\% C.L. upper cross section 
limits for gluino-like pair production processes with produced particles decaying 
into $b\bar{b}\neut$ final states. The cross section upper limits versus the gluino and neutralino mass are also given in Figure~\ref{fig:GbbMaxAll}. 

\begin{figure}[tb]
  \begin{center}
      \includegraphics[width=0.58\columnwidth]{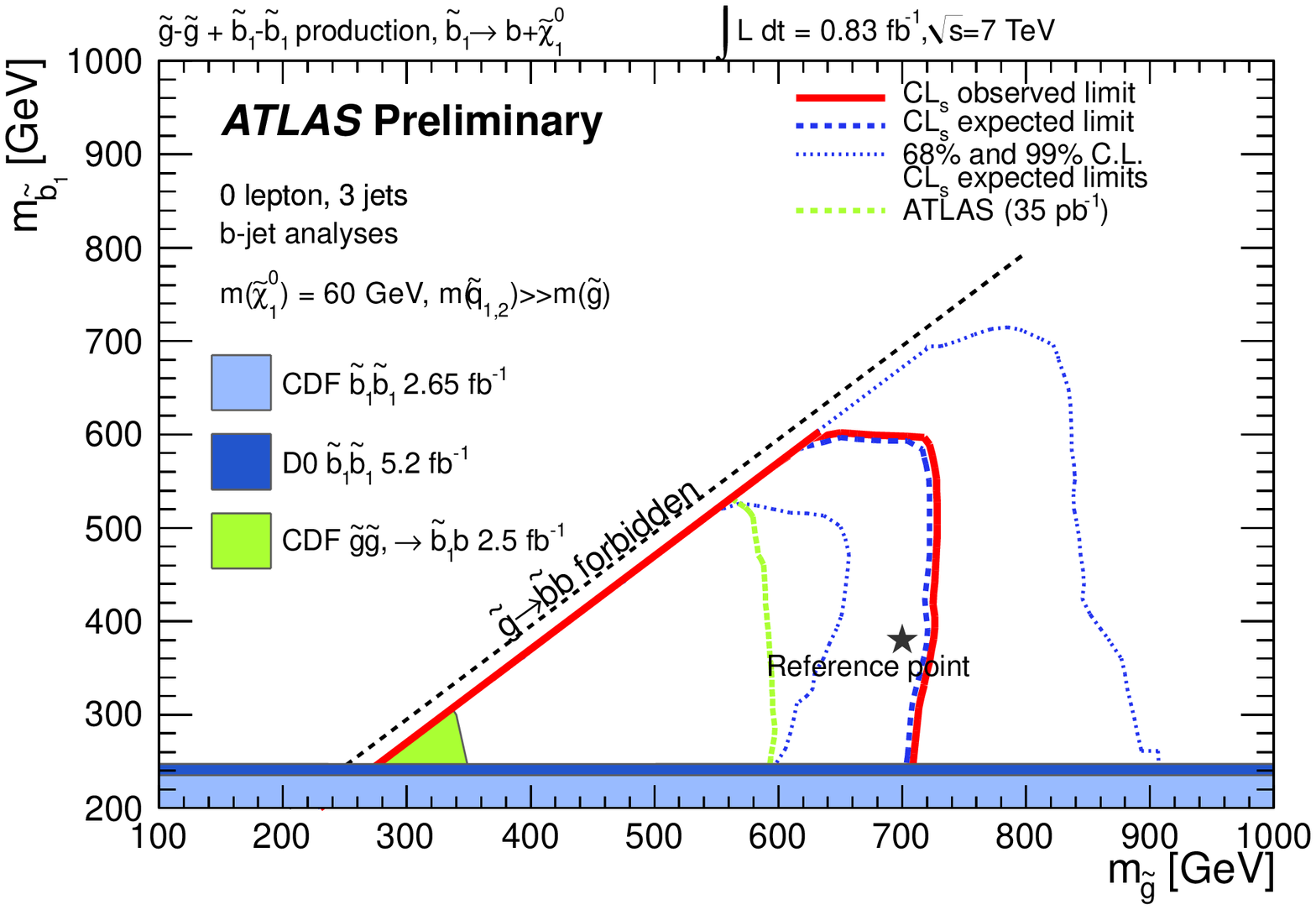}
    \end{center}
  \caption{Observed and expected 95\% C.L. exclusion limits in the ($m^{}_{\gl},m^{}_{\bone}$) plane. 
  Also shown are the 68\% and 99\% C.L. expected exclusion curves.
  For each point in the plot, the signal region selection providing 
  the best expected limit is chosen. The neutralino mass is set to  60~GeV. 
  The result is compared to previous results from ATLAS and CDF searches which assume the same gluino-sbottom 
  decays hypotheses. Exclusion limits from the CDF and D0 experiments on direct sbottom pair production are 
  also shown. }
  \label{fig:sb_gl_obs}
\end{figure}
\begin{figure}[h!]
 \begin{center}
     \includegraphics[width=0.68\columnwidth]{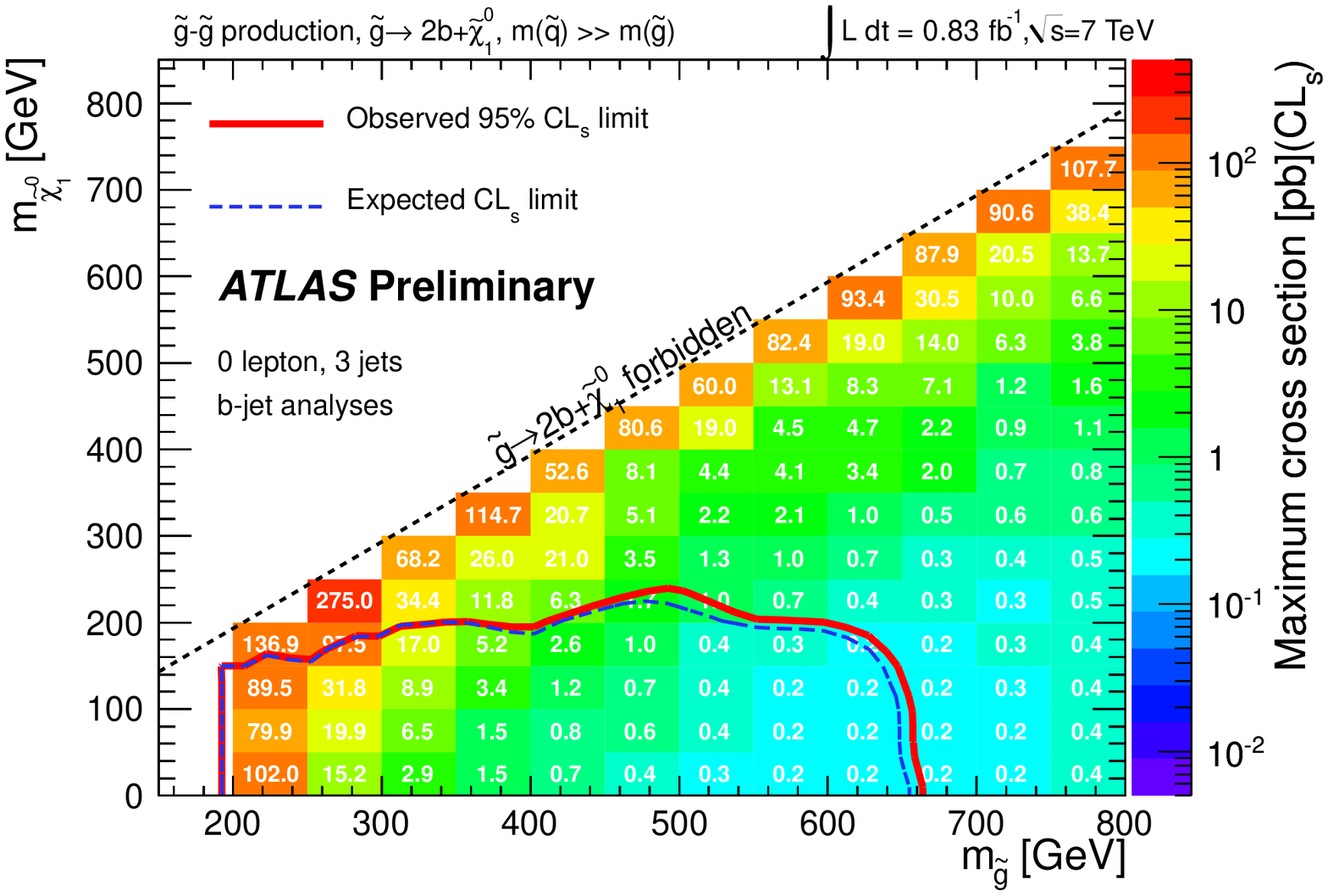} 
   \end{center}
 \caption{95\% C.L. upper cross section limits in pb and observed and expected limit contours  
  in the ($m^{}_{\gl},m^{}_{\neut}$) plane for gluino masses above 200 GeV. For each scenario, 
  the signal region selection providing the best expected limit is chosen.}
  \label{fig:GbbMaxAll}
\end{figure}

%%%%%%%%%%%%%%%%%%%%%%%%%%%%%%%%%%
\section{Conclusions}
An update on the search for supersymmetry in final states with missing transverse 
momentum, $b$-jet candidates and no isolated leptons in proton-proton collisions 
at 7~TeV is presented. The results are based on data corresponding 
to an integrated luminosity of \lumi  ~\ifb\ collected during 2011 by the ATLAS experiment at the LHC. Events with at least three energetic jets, large 
\met and at least one \btagged jet are selected in four signal regions based on the number of \btagged jets ($\geq$1 or $\geq$2 \bjets) and on the value 
of effective mass ($>$500 or $>$700 GeV).  The dominant Standard Model backgrounds are estimated from Monte Carlo simulation and are validated with data.

No excess above the expectation from Standard Model processes is found. The 
results are used to exclude parameter regions in various $R$-parity conserving SUSY models. 
Under the assumption that the lightest squark $\tilde{b_1}$ is produced via 
gluino-mediated processes or direct pair production and decays exclusively via 
$\tilde{b_1} \rightarrow b\neut$, and that $m_{\neut}$ = 60 GeV, gluino masses below 720~GeV are excluded 
with 95\% C.L. for sbottom masses up to 600~GeV using the $CL_s$ approach. 
This extends the previous (35 \ipb) ATLAS limits on gluino masses in the same scenario by 
about 130 GeV. Results are also interpreted in simplified models, 
where gluinos decay into heavy flavor final states ($b\bar{b}\neut$) via an 
off-shell sbottom. In these scenarios exclusion limits in the 
($m^{}_{\gl},m^{}_{\neut}$) plane are derived and $\neut$ masses below 200-250~GeV are excluded for 
gluino masses below 660~GeV, if $\Delta M (\gl-\neut)>$100 GeV. 
95\% C.L. upper cross section 
limits for gluino-like pair production processes with produced particles decaying 
into $b\bar{b}\neut$ final states are also given.

%%%%%%%%%%%%%%%%%%%%%%%%%%%%%%%%%%
\begin{acknowledgments}
This proceeding is adapted from an ATLAS conference note \cite{ATLAS-CONF-2011-098}, and as such, includes text written by several members of the ATLAS collaboration.
\end{acknowledgments}

\bigskip % extra skip inserted
%\bibliography{dpf2011_bbutler}
%\begin{thebibliography}{9}   % Use for  1-9  references
%\begin{thebibliography}{99} % Use for 10-99 references

%\bibitem{charm07}   http://www.lepp.cornell.edu/charm07/

%\bibitem{templates-ref} http://www.slac.stanford.edu/econf/editors/eprint-template/instructions.html

%\end{thebibliography}

\end{document}